\documentclass[aps,prb,twocolumn,nobibnotes]{revtex4}  
\usepackage{graphicx}

\begin{document}

\title{Measurement of one-photon and two-photon wavepackets\\ in spontaneous parametric down-conversion}

\author{Yoon-Ho Kim}\email{yokim@umbc.edu; kimy@ornl.gov}

\affiliation{Center for Engineering Science Advanced Research\\ Computer Science \& Mathematics Division\\ Oak Ridge National Laboratory\\ Oak Ridge, Tennessee 37831}

\date[]{to appear in J. Opt. Soc. Am. B}

\begin{abstract}
One-photon and two-photon wavepackets of entangled two-photon states in spontaneous parametric down-conversion (SPDC) fields are calculated and measured experimentally. For type-II SPDC, measured one-photon and two-photon wavepackets agree well with theory. For type-I SPDC, the measured one-photon wavepacket agree with the theory. However, the two-photon wavepacket is much bigger than the expected value and the visibility of interference is low. We identify the sources of this discrepancy as the spatial filtering of the two-photon bandwidth and non-pair detection events caused by the detector apertures and the tuning curve characteristics of the type-I SPDC. 
\end{abstract}

\maketitle

\section{Introduction}

The two-photon state generated via spontaneous parametric down-conversion (SPDC) is one of the most well-known examples of two-particle entangled states. The SPDC process can be briefly explained as a spontaneous splitting or decay of a pump photon into a pair of daughter photons (typically called signal and idler photons) in a nonlinear optical crystal \cite{klyshko}. This spontaneous decay or splitting only occurs when energies and momentum of the interacting photons satisfy the conservation condition, which is known as the phase matching condition. Signal and idler photons have the same polarization in type-I phase matching and have orthogonal polarization in type-II phase matching.

If the phase matching is perfect, assuming monochromatic plane wave pump, it is not hard to see that the state of SPDC should be written as
\begin{eqnarray}\label{eq:perfect}
|\Psi\rangle &=& \sum_{s,i} \delta(\omega_s+\omega_i-\omega_p) \delta(\mathbf{k}_s+\mathbf{k}_i-\mathbf{k}_p)\nonumber\\
&&\times a_s^\dagger(\omega(\mathbf{k}_s)) a_i^\dagger(\omega_i(\mathbf{k}_i))
|0\rangle,
\end{eqnarray}
where $\omega_j$, $\mathbf{k}_j$ ($j=s,i,p$) are the frequency and the wave vectors of the signal, idler, and the pump, respectively, and $a_s^\dagger(\omega(\mathbf{k}_s))$ is the creation operator for the signal photon. The $\delta$ functions in state (\ref{eq:perfect}) ensure that the signal and the idler photons satisfy the phase matching condition, i.e., there is only one energy or wave vector for the idler photon corresponding to a given energy or wave vector for the signal photon. In other words, the signal-idler photon pair is perfectly entangled in energy and momentum. Such a perfectly entangled two-photon state (in energy and momentum) naturaly has an infinite two-photon coherence time.

In reality, such a strict one-to-one correspondence does not happen, because
perfect phase matching can never occur, most notably, due to pump beam
divergence, limited pump beam size, and limited thickness of the nonlinear
crystal \cite{hong,rubin}. Therefore, the above delta functions
should be replaced by two-photon spectral functions that are sharply peaked
around $\omega_s=\omega_p-\omega_i$ and $\mathbf{k}_s=\mathbf{k}_p-\mathbf{k}_i$ and have some bandwidths. This means that given an energy or momentum of the signal photon, there are some ranges of energies and momenta available for the idler photon: the entanglement between the photon pair is less-than-perfect. As a result, the two-photon state of SPDC has finite coherence time and the shape of the correlation function is determined by the two-photon spectral functions, which depend on the types of phase matching. (Note that monochromatic pumping condition is assumed in this paper.) We may then define the one-photon and the two-photon wavepackets of the SPDC as the envelope of the first-order interference observed in the single-detector count rate and the envelope of the second-order interference observed in the coincidence count between two detectors, respectively.

In this paper, we present theoretical calculation and experimental measurements of  the one-photon and the two-photon wavepackets of SPDC  for both type-I and type-II phase matching conditions. First, in section
\ref{sec:coh1}, we calculate the first-order ($G^{(1)}(\tau)$) and the second-order  ($G^{(2)}(\tau)$) correlation functions for the quantum state of SPDC. We then calculate, in section \ref{sec:coh2},  (i) the one-photon wavepacket: the envelope of first-order interference fringe due to a Michelson interferometer  and (ii) the two-photon wavepacket: the envelope of second-order interference due to a Shih-Alley/Hong-Ou-Mandel interferometer using the SPDC state as the input \cite{shih,shih1,hom,hom1,kwiat1,shih2,shih3}. 
It is found that the two-photon wavepacket measured this way is not related to the second-order correlation function $G^{(2)(\tau)}$ of the state  but related to the
first-order correlation function ($G^{(1)}(\tau)$) of the state \cite{burlakov}. These predictions are experimentally tested in section \ref{sec:exp} using type-II and type-I SPDC.
It turns out that, the experimental results for the type-I SPDC case do not quite
agree with the predictions in a realistic experimental setup. Possible reasons for such deviations are discussed.

\section{Correlation functions of SPDC}\label{sec:coh1}

The quantum state of SPDC can be calculated using first-order perturbation
theory \cite{rubin},
$$
|\psi\rangle = -\frac{i}{\hbar}\int_{-\infty}^\infty dt \mathcal{H}|0\rangle.
$$
$\mathcal{H}$ is the interaction Hamiltonian which takes the form
$$
\mathcal{H}= \epsilon_0 \chi^{(2)}\int_V d^3 \mathbf{r} E_p^{(+)} E_s^{(-)}
E_i^{(-)} + h.c.,
$$
where $E_p^{(+)}$ is the pump laser field which is considered monochromatic
(cw) and classical. Assuming that it is propagating in $z$ direction and has
frequency $\Omega_p$, $E_p^{(+)}=\mathcal{E}_p \, \exp[i(k_p z - \Omega_p t)]$.
$E_j^{(-)}$, ($j=s,i$), is the quantized field operator for the signal and the
idler photons. Assuming also that $E_j^{(-)}$ is propagating in  $z$ direction
and has central frequency $\omega_j$, it can be written as $E_j^{(-)}=\int_V \,
d^3 \mathbf{k} \, \mathcal{E}_j \, a^\dagger_j(\omega_j) \exp[-i(k_j z -
\omega_j t)]$. The state of SPDC can then be calculated as \cite{rubin},
\begin{equation}\label{eq:eq0}
|\psi\rangle = \int d\omega_{s} \, d\omega_{i} \,\mathrm{sinc}\left(\frac{\Delta L}{2}\right) e^{-i\frac{\Delta L}{2}} a^{\dagger}_{s}(\omega_{s}) a^{\dagger}_{i}(\omega_{i})|0\rangle,
\end{equation}
where $\Delta = \mathbf{k}_{p}(\Omega_{p})-\mathbf{k}_{s}(\omega_{s})-\mathbf{k}_{i}(\omega_{i})$ and $L$ is the thickness of the nonlinear crystal. (We have ignored the normalization constant for simplicity.)

Since the condition $\omega_{s}+\omega_{i}=\Omega_{p}$ has to be satisfied at all times, we can simplify the above equation further by introducing the detuning frequency $\nu=\omega_s-\Omega$ or,  equivalently, $\nu=\omega_i +\Omega$ where $\Omega=\Omega_{p}/2$. We therefore obtain
\begin{equation}\label{eq:state}
|\psi\rangle = \int_{-\infty}^\infty d\nu \, T(\nu) \, 
a_s^\dagger(\Omega+\nu)\, a_i^\dagger(\Omega-\nu)|0\rangle,
\end{equation}
where $T(\nu)=S(\nu) P(\nu)$.  $S(\nu)$ is the joint spectral function for the signal and the idler photons which determines the coherence properties of the state and $P(\nu)$ is the frequency dependent phase term. Note also that this expression clearly shows the frequency anti-correlated feature of the signal and the idler photons.

The specific forms of $S(\nu)$ depend on the phase matching condition used in
an experiment and are well-known \cite{klyshko,rubin,valencia}. For type-II
SPDC,
$$
S(\nu) = \textrm{sinc}\left(\frac{\nu D L}{2}\right),
$$
where the group velocity difference (in the crystal) $D =  dK_i/d\Omega_i -
dK_s/d\Omega_s$ and $L$ is the thickness of the nonlinear crystal. In type-I
SPDC, it is given as
$$
S(\nu) = \textrm{sinc}\left(\frac{\nu^2 D^{''} L}{2}\right),
$$
where the group velocity dispersion (in the crystal) $D^{''} = d^2
K/d\Omega^2$. (Note $K_i = K_s$ for type-I SPDC.)

Let us now introduce the density operator as it is convenient to use the
reduced density operator to calculate the first-order correlation function of
the state. Since the two-photon state (\ref{eq:state}) is a pure state, the
density operator for the two-photon state is simply given as $\hat{\rho} =
|\psi\rangle\langle\psi|$. To obtain the density operator for the signal
photon, we need to perform a partial trace of the two-photon density operator
\cite{strekalov},
\begin{eqnarray}\label{eq:den}
\hat{\rho}_s &=& tr_i[\hat{\rho}]\nonumber\\
&=&\int_{-\infty}^{\infty} d\nu \, |S(\nu)|^2 \,
a_s^\dagger(\Omega+\nu)|0\rangle\langle0|a_s(\Omega+\nu).
\end{eqnarray}

First-order correlation function of the state can be calculated using the
reduced density operator obtained in Eq.~(\ref{eq:den}). For stationary fields,
the first-order correlation function can be written as \cite{mandel}
\begin{equation}\label{g1}
G^{(1)}(\tau) = tr[\hat{\rho}_s E^{(-)}_s(t)E^{(+)}_s(t+\tau)],
\end{equation}
where $E^{(-)}_s(t)=\int_0^\infty d\omega \, a^\dagger_s(\omega) \exp[i \omega
t]$. The first-order correlation function of the signal photon can then be
calculated as
\begin{equation}
G^{(1)}(\tau) = \int_0^\infty d\omega |S(\omega-\Omega)|^2 e^{-i \omega \tau},
\end{equation}
where $\omega-\Omega = \nu$. As we can see clearly, first-order correlation
function of the signal photon is simply a Fourier transform of the power
spectrum of the signal photon.

Second order correlation function can be calculated rather simply by using
state (\ref{eq:state}),
\begin{equation}\label{g2}
G^{(2)}(\tau) = |\langle0|E_2^{(+)}(t+\tau)E_1^{(+)}(t)|\psi\rangle|^2,
\end{equation}
where $E_1^{(+)}(t+\tau)=\int_0^\infty d\omega_s \, a_s(\omega_s) \, \exp[-i
\omega_s (t+\tau)]$ and $E_2^{(+)}(t)=\int_0^\infty d\omega_i \, a_i(\omega_i)
\, \exp[-i \omega_i t]$. By using $\omega_s = \Omega + \nu$ and $\omega_i =
\Omega-\nu$, it is straightforward to obtain
\begin{equation}
G^{(2)}(\tau)=\left|\int_{-\infty}^{\infty} S(\nu) e^{-i \nu \tau} \right|^2.
\end{equation}

Note that $G^{(1)}(\tau)$ and $G^{(2)}(\tau)$ can have quite different shapes
even though they are associated with the same $S(\nu)$. For example,
$G^{(1)}(\tau)$ does not get affected by the introduction of group velocity
dispersion between the source and the detector, but $G^{(2)}(\tau)$ gets
broadened by it \cite{valencia}. It is because any dispersion introduced in
$E_s^{(-)}(t)$ simply cancels when calculating $G^{(1)}(\tau)$. In the case of
$G^{(2)}(\tau)$, this cancellation does not happen because two different fields
are involved \cite{note}.

\section{One-photon and two-photon wavepackets}\label{sec:coh2}

We have so far calculated first- and second-order correlation functions of the
quantum state of SPDC. In this section, we study how these correlation
functions are actually linked to one-photon and two-photon wavepackets in simple
interference experiments.

For one-photon wavepacket measurement, we consider the output of a simple Michelson interferometer, in which either the signal or the idler photons are the input. For the two-photon wavepacket measurement, we consider a well-known
Shih-Alley/Hong-Ou-Mandel interferometer setup, in which the signal-idler
photon pair is made to interfere at a beamsplitter and the coincidence counts
between detectors, which are placed at the output ports of the beamsplitter,
are measured \cite{shih,shih1,hom,hom1,kwiat1,shih2,shih3}.

Let us first calculate the single count rates at the output port of a Michelson
interferometer when the signal photon of SPDC is the input. In this case, the single
count rate is proportional to $R_s$,
\begin{equation}
R_s = tr[\hat{\rho}_s E^{(-)}(t) E^{(+)}(t)],
\end{equation}
where the reduced density operator for the signal photon, $\hat{\rho}_s$, is
given in Eq.~(\ref{eq:den}), $E^{(+)}(t)= \int \left\{ a(\omega) \exp[-i\omega
t] + a(\omega) \exp[-i\omega(t+\tau)]\right\}d\omega$, and $\tau$ is the delay
between the two arms of the interferometer. It is then easy to show that
$$
R_s=\int_0^\infty |S(\omega-\Omega)|^2 \left\{ 1+\cos(\omega \tau) \right\},
$$
which can be re-written as
\begin{equation}
R_s = \frac{1}{2}\left\{ 1 + g^{(1)}(\tau)\cos(\Omega \tau)\right\},
\end{equation}
where $g^{(1)}(\tau) = |G^{(1)}(\tau)|/|G^{(1)}(0)|$. Therefore, the envelope
of the interference fringe or the one-photon wavepacket observed at the output of the Michelson interferometer directly corresponds to the first-order correlation function
$G^{(1)}(\tau)$.

\begin{figure}[t]
\centerline{\scalebox{.65}{\includegraphics{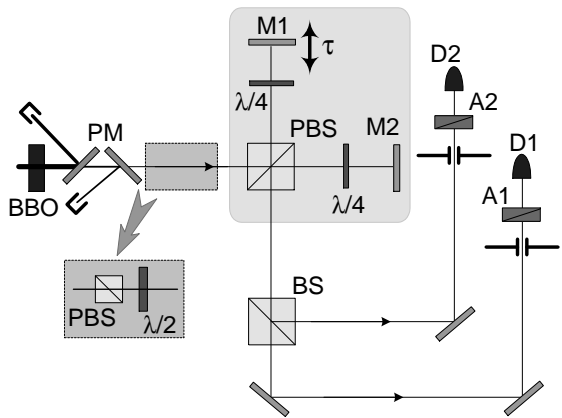}}}
\caption{\label{fig:type2fig}Experiment using collinear type-II SPDC.
$\lambda/2$ plate is oriented at $22.5^\circ$ and $\lambda/4$ plates are
oriented at $45^\circ$. Polarizer (PBS) and $\lambda/2$ plate set, shown in the
inset, is inserted only when first-order interference is measured. Shaded area
containing PBS, $\lambda/4$ plates, and mirrors is equivalent to commonly used
quartz polarization delay.}
\end{figure}

For a two-photon wavepacket, we need to  calculate the coincidence count rates for a Shih-Alley/Hong-Ou-Mandel interferometric setup \cite{shih,shih1,hom,hom1,kwiat1,shih2,shih3}. Consider the following setup: signal and idler photons are generated at the crystal, propagate at different directions, reflect off at mirrors, and made to interfere at a beamsplitter. If both photons have the same polarization,
polarization of one of the photons is rotated by $90^\circ$ before reaching the
beamsplitter. The delay between the two paths are $\tau$. A detector package
which consists of a single photon detector and a polarization analyzer is
placed at each output port of the beamsplitter and the coincidence counts
between the two detectors are recorded (See experimental setup shown in Fig.~\ref{fig:type2fig} and
Fig.~\ref{fig:type1fig}). The coincidence count rate is then
proportional to $R_c$,
\begin{equation}\label{eq:rc1}
R_c = \int | \langle 0 | E_2^{(+)}(t_2) E_1^{(+)}(t_1)|\psi\rangle |^2 dt_1
dt_2.
\end{equation}
The quantized electric fields $E_2^{(+)}(t_2)$ and  $E_1^{(+)}(t_1)$ at the
detectors $D_1$ and $D_2$ can be written as
\begin{widetext}
\begin{eqnarray}
E_2^{(+)}(t_2) &=& -i\sin\theta_2 \int d\nu \, a_s(\Omega+\nu)
e^{-i(\Omega+\nu)t_2} + \cos\theta_2 \int d\nu \, a_i(\Omega-\nu)
e^{-i(\Omega-\nu)(t_2+\tau)},\nonumber\\
E_1^{(+)}(t_1) &=& -\sin\theta_1 \int d\nu \, a_s(\Omega+\nu)
e^{-i(\Omega+\nu)t_1} + i\cos\theta_1 \int d\nu \, a_i(\Omega-\nu)
e^{-i(\Omega-\nu)(t_1+\tau)},\nonumber
\end{eqnarray}
\end{widetext}
where $\theta_1$ and $\theta_2$ are the angles of the polarization analyzers
placed before the detectors, $\tau$ is the delay introduced between the two
arms of the interferometer, $t_1$ and $t_2$ are the times at which the
detectors $D_1$ and $D_2$ click, and the phase factor $i$ comes from the
reflection at the beamsplitter. To trace the envelope of the interference,
$\theta_1$ and $\theta_2$ values should be chosen so that the maximum and minimum
of the interference can be observed for a certain value of $\tau$. These values
are $\theta_1=\theta_2=45^\circ$ for the interference minima and
$\theta_1=-\theta_2=45^\circ$ for the interference maxima
\cite{shih,shih1,hom,hom1,kwiat1,shih2,shih3}.

Further evaluating Eq.~(\ref{eq:rc1}) for $\theta_1=\theta_2=45^\circ$, we get
\begin{eqnarray}
R_c &=& \int dt_+ dt_- \int d\nu d\nu' S(\nu) S(\nu')
e^{i(\nu-\nu')\tau}\nonumber\\ && \times \sin(\nu t_-)\sin(\nu' t_-),\nonumber\\
&\approx& \int d\nu |S(\nu)|^2 - \int d\nu |S(\nu)|^2 e^{-i 2\nu\tau},\nonumber\\
&=& 1 - g^{(1)}(2\tau),\nonumber
\end{eqnarray}
where $t_-=t_2-t_1$, $t_+=t_1+t_2$. In approximating the above equation, we've
used the fact that $S(\nu)$ is an even function and $\nu$ is in optical
frequency. The above equation can be re-written as
\begin{equation}\label{eq:rc2}
R_c = \frac{1}{2} \left\{ 1 \pm g^{(1)}(2\tau) \right\},
\end{equation}
where $-$ sign is for $\theta_1=\theta_2=45^\circ$ and $+$ sign is for
$\theta_1=-\theta_2=45^\circ$.

It is interesting to note that the two-photon wavepacket, eq.~(\ref{eq:rc2}), does not contain the second-order correlation function $G^{(2)}(\tau)$. This result has interesting
implications: (i) $R_c$ has the same envelope as $R_s$ except that the $R_c$
envelope is half of the $R_s$ envelope, (ii) any dispersion element introduced
in a Shih-Alley/Hong-Ou-Mandel interferometer cannot affect the shape of the
interference envelope since $g^{(1)}(\tau)$ is not affected by group velocity
dispersion \cite{note2}.

\section{Experiment}\label{sec:exp}

In this section, we describe two experiments which are designed to test the
predictions made in section \ref{sec:coh2}. For both type-I and type-II SPDC
experiments, the pump laser was a argon ion laser operating at 351.1 nm.
Coincidence counts were measured using a time-to-amplitude converter (TAC) and
multi-channel analyzer (MCA) set. The coincidence window used for second-order
interference measurement was 3 nsec.

\begin{figure}[t]
\centerline{\scalebox{.65}{\includegraphics{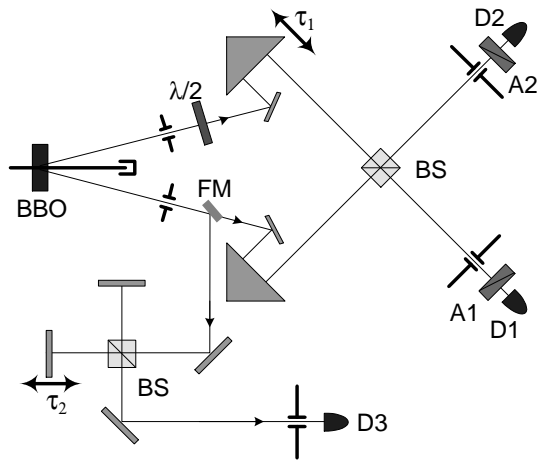}}}
\caption{\label{fig:type1fig}Experimental setup using non-collinear type-I
SPDC. $\lambda/2$ plate rotates the polarization of the signal photon from
horizontal to vertical. FM is a flipper mirror.}
\end{figure}

Let us first describe the wavepacket measurement of type-II SPDC. The
experimental setup can be seen in Fig.~\ref{fig:type2fig}. A 2 mm thick type-II
BBO crystal was pumped by a 351.1 nm laser beam generating 702.2 nm collinear
type-II SPDC photons. The residual pump beam was removed by two pump reflecting
mirrors (PM). Instead of using usual quartz delay line for introducing fine
delay between horizontal and vertically polarized photons \cite{shih2,shih3}, a set of PBS, $\lambda/4$ plates (oriented at $45^\circ$), and mirrors (shown in the shaded area) was used. The inset containing a PBS and a $\lambda/2$ plate (oriented at $22.5^\circ$) was used to remove vertically polarized photons when measuring
first-order interference. The delay between the two arms of the interferometer
was introduced by moving mirror M1 with an encoder driver. Photons were finally
detected with detector packages which consist of a single-photon counting
module and a polarization analyzer. The distance from the BBO crystal to the
detector was approximately 218 cm and all apertures used in this experiment
were about 3 mm in diameter.

\begin{figure}[t]
\centerline{\scalebox{.73}{\includegraphics{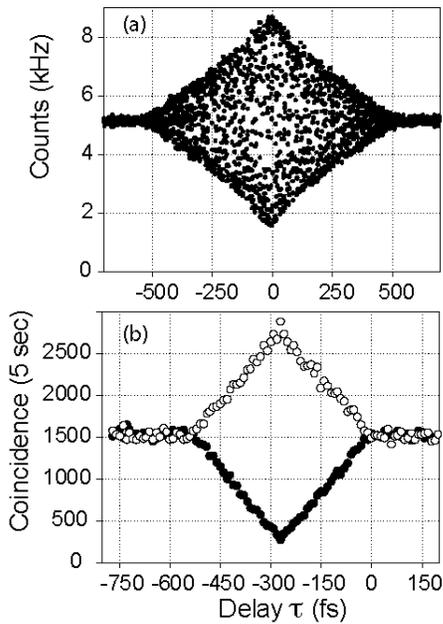}}}
\caption{\label{fig:dataII}Experimental data for type-II SPDC. (a) First-order interference. (b) Second-order interference.  Solid circles are for
$\theta_1=\theta_2=45^\circ$ and empty circles are for
$\theta_1=-\theta_2=45^\circ$, where $\theta_1$ and $\theta_2$ are analyzer (A1
and A2) angles. Peak-dip visibility is about 84 \%. Coincidence peak or dip occurs when the e-polarized photons are delayed by $D\times L/2 \approx 247$ fs with respect to the o-polarized photons before reaching the beamsplitter BS.}
\end{figure}

\begin{figure}[t]
\centerline{\scalebox{0.5}{\includegraphics{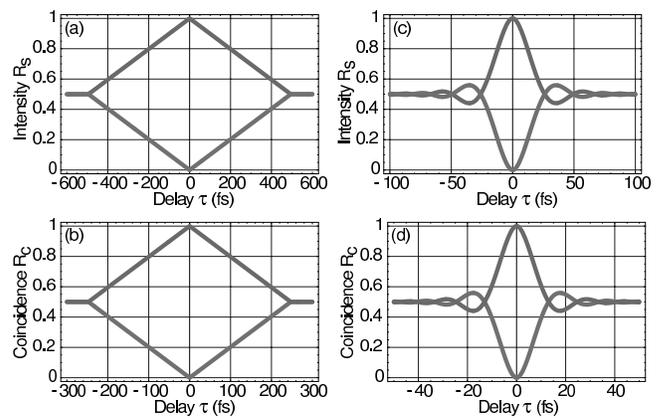}}}
\caption{\label{fig:theory}Calculated first- and second-order interference patterns  for type-II SPDC, (a) and (b), and type-I SPDC, (c) and (d).  Only the fringe envelopes are shown for the first-order interference $R_s$. It is clear that $R_{s}$ and $R_{c}$ have the same envelope shapes. However, the width of the coincidence envelope is half of that of first-order interference ($R_s$). The plots are calculated for the following parameters: BBO crystal with 2 mm thickness, 351.1 nm pump wavelength, and 702.2 nm SPDC central wavelength. For (b), the delay is shifted by $D\times L/2 \approx 247$ fs for easy comparison with (a). }
\end{figure}

Fig.~\ref{fig:dataII} shows the experimental data for the type-II SPDC
experiment. Fig.~\ref{fig:dataII}(a) shows first-order interference of the
horizontal polarized photon. As discussed above, vertical polarized photons are
removed with a PBS and a $\lambda/2$ plate is used to rotate the polarization
direction to $45^\circ$. For this measurement, we used a 20 nm FWHM filter to
suppress the white-light interference which occurs around $\tau=0$
\cite{strekalov}. The observed triangular one-photon wavepacket agrees well with the theoretical prediction shown in Fig.~\ref{fig:theory}(a).

To measure the two-photon wavepacket, we first removed the PBS-$\lambda/2$
plate set used for first-order interference measurement. The e-ray of the
crystal (vertically polarized photons) could then be delayed with respect to
the o-ray by moving M1. For this measurement, only uv-cut off filters (cut-off
at 550 nm) were used to suppress any residual pump noise.
Fig.~\ref{fig:dataII}(b) shows typical triangular two-photon wavepacket
observed in coincidence counts in which the dip or peak occurs when the e-polarized photons are delayed by $D\times L/2\approx247$fs with respect to the o-polarized photons before reaching the beamsplitter BS \cite{shih2,shih3}. Again, the observed triangular two-photon wavepacket agrees well with the theoretical prediction shown in Fig.~\ref{fig:theory}(b).

As we have predicted in the previous section, the one-photon and the two-photon wavepackets have the same shapes and the one-photon wavepacket is twice bigger than the two-photon wavepacet.  Although we here have used collinear type-II SPDC for one-photon and two-photon wavepacket calculation and measurements, recent experimental results confirm that non-collinear type-II SPDC gives the same result for the two-photon wavepacket measurement \cite{kg1,kg}. Finally, we note that the
power spectrum function, in type-II SPDC, includes parameters of both the
signal and the idler photons (in $D =  dK_i/d\Omega_i -
dK_s/d\Omega_s$) even though only one of them is actually
measured \cite{strekalov}.

Let us now discuss the measurement of one-photon and two-photon wavepackets for type-I SPDC. The experimental setup can be seen
in Fig.~\ref{fig:type1fig}. The pump laser beam was centered at 351.1 nm and
702.2 nm centered SPDC photons were generated from a 2 mm thick type-I BBO
crystal. The propagation angle of the signal-idler photon pair was about $\pm
3^\circ$ with respect to the pump beam propagation direction. A $\lambda/2$
plate was used to rotate the signal photon's polarization from horizontal
polarization to vertical polarization. The signal-idler photons were then made
to interfere at a beamsplitter and the delay $\tau_1$ was varied by an encoder
driver driven trombone prism. Detectors D1 and D2 placed at the output ports of
the beamsplitter were used for second-order interference measurement. For
first-order interference measurement, a flipper mirror (FM) was used to direct
the idler photon to the secondary Michelson interferometer. Interference was
measured by detector D3 as a function of the arm length difference $\tau_2$.
The crystal to the D3 distance was about 200 cm and to D2-D1 was about 280 cm.
As before, all apertures used in this experiment were about 3 mm in diameter.

\begin{figure*}[t]
\centerline{\scalebox{0.65}{\includegraphics{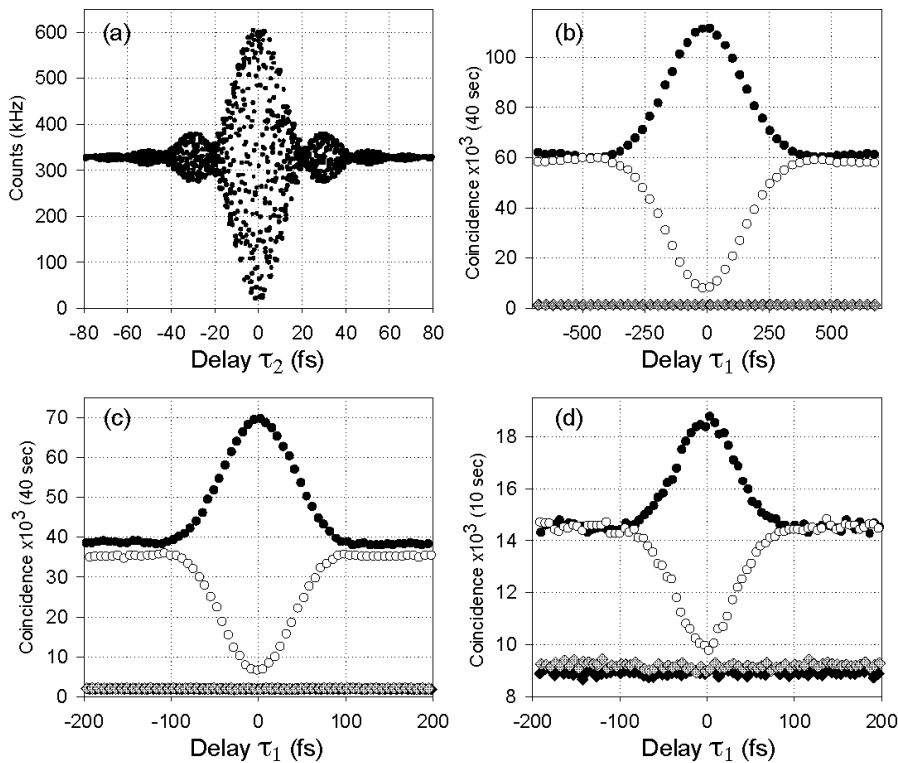}}}
\caption{\label{fig:dataI}Experimental data for type-I SPDC experiment. (a)
First-order interference. Only UV-cutoff filter (cutoff at 550 nm) is used to
suppress the pump noise. The visibility is about 92\%. (b), (c), and (d) show
second-order interference measurements. Diamond data points show the level of
accidental coincidence. Solid data points are for $\theta_1=-\theta_2=45^\circ$
and empty circles are for $\theta_1=\theta_2=45^\circ$. Spectral filters used
for (b), (c), and (d), are 3 nm FWHM, 20 nm FWHM, and 80 nm FWHM,
respectively.}
\end{figure*}

For one-photon wavepacket measurement, we used the flipper mirror (FM) to direct the idler photon to the secondary Michelson interferometer as shown in Fig.~\ref{fig:type1fig}. First-order interference was observed by moving one of the mirror by an encoder driver and to reduce any residual pump noise, a uv-cutoff filter was used in front of the detector. The experimental data for this measurement can be seen in Fig.~\ref{fig:dataI}(a) and  the envelope of first-order
interference or one-photon wavepacket closely follows the predicted curve shown in
Fig.~\ref{fig:theory}(c). This confirms that the power spectrum of either the
signal or the idler photon of type-I SPDC is indeed given by
$|S(\omega-\Omega)|^2$. 

For two-photon wavepacket measurement, we removed the flipper mirror (FM)
and let the signal-idler photons interfere at the beamsplitter. The coincidence
counts were measured as a function of both the delay $\tau_1$ and the analyzer
angles. We also measured the true coincidence counts as well as accidental
coincidence counts by integrating additional 3 nsec window which was located
about 10 nsec away from the true coincidence peak in the MCA spectrum.

We first measured the two-photon wavepacket with 3 nm FWHM spectral filters.
The data for this experiment can be seen in Fig.~\ref{fig:dataI}(b). The
visibility for this measurement is quite high ($\sim 87$\%), however, the
two-photon wavepacket has quite different shape than the one-photon wavepacket, mostly due to narrow-band filtering of the SPDC
photons by the 3 nm filters. We can roughly estimate the contribution of the
spectral filters to the broadened coherence time by using $\tau_c \sim
\lambda^2/(c \Delta \lambda) \approx 550$ fsec. Considering that the filters do
not necessarily have perfect Gaussian shape transmission curve and they may
have different FWHM values than the specified values, this rough estimation
gives a pretty good idea on the origin of the broadened two-photon wavepacket. Note
also that the level of accidental coincidence is nearly negligible.

The same measurement was repeated with two more sets of spectral filters: 20 nm
FWHM and 80 nm FWHM. With 20 nm filters, see Fig.~\ref{fig:dataI}(c), we still
observe quite good visibility of 83\%. Notice that the level of accidental
contribution has risen slightly and the two-photon wavepacket is
now narrower. By using the simple picture again, we estimate $\tau_c \sim
\lambda^2/(c \Delta \lambda) \approx 82$ fsec. This value is quite
close to the observed two-photon wavepacket shown in Fig.~\ref{fig:dataI}(c) and it  means that we are still observing a spectrally filtered, by the spectral filters, two-photon wavepacket. 

Finally, 80 nm FWHM filters were used for the same measurement, see
Fig.~\ref{fig:dataI}(d). We find that the raw visibility has now dropped
to 32\% with almost no change in the two-photon wavepacket.
According to Fig.~\ref{fig:theory}(d), the unfiltered two-photon wavepacket should be about 15 fsec in FWHM. Note also that the contribution of accidental coincidence is now significant, unlike Fig.~\ref{fig:dataI}(b) and Fig.~\ref{fig:dataI}(c).

This rather unexpected behavior of the two-photon wavepacket with broadband spectral filters can be understood as follows. It is well-known that type-I SPDC in general has much bigger bandwidth than type-II SPDC. Especially, for the case considered in this paper, the calculated FWHM of the type-I SPDC spectrum is more than 80 nm which is much bigger than roughly 3 nm calculated FWHM bandwidth of type-II SPDC. This calculation actually agrees quite well with the observed first-order interference shown in Fig.~\ref{fig:dataII}(a) and Fig.~\ref{fig:dataI}(a).
Based on this observation alone, it may seem, at first, that the bandwidth of
type-I SPDC is not limited at all at the detectors. To see what really is
happening, however, it is necessary to consider the type-I SPDC tuning curve, which  shows how SPDC spectrum is distributed as functions of propagation angles and wavelengths.

\begin{figure}[t]
\centerline{\scalebox{0.5}{\includegraphics{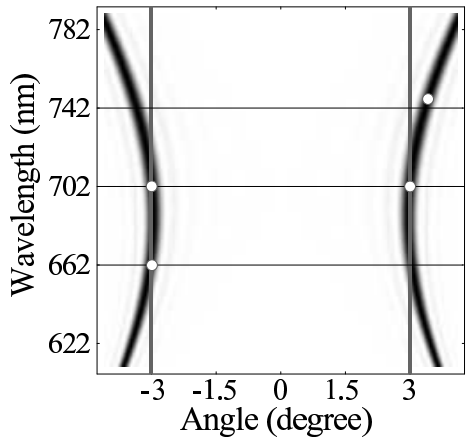}}}
\caption{\label{fig:tuning}Tuning curve for non-collinear type-I SPDC used in
this experiment. Two vertical bars located at $\pm 3^\circ$ represent the
angles defined by the apertures. Left (right) curve shows the angle-spectrum
distribution of the signal (idler) photons. See text for details.}
\end{figure}

Fig.~\ref{fig:tuning} shows the tuning curve of non-collinear type-I SPDC used
in this experiment. The left curve shows the angle-spectrum distribution for
the signal photons and the right curve shows the same for the idler photons.
Note that the signal and the idler photons have the same angle-spectrum
distribution as both signal and idler photons have the same polarization. Two
vertical bars represent the angles defined by the apertures used in the
experiment.

For one-photon wavepacket measurement, only the signal or the idler photons
are measured. If is clear from Fig.~\ref{fig:tuning} that the signal or the
idler photons indeed have quite broad bandwidth (more than 80 nm) even if we
only consider the small angle defined by the aperture. It is because the slope
of the tuning curve for type-I SPDC is not steep, unlike type-II SPDC. For
second-order interference measurement, however, we need to consider
signal-idler photon pair detections and the fact that their frequencies are
anti-correlated, i.e., $\omega_s = \Omega+\nu$ and $\omega_i=\Omega-\nu$. If
spectral filers have narrow bandwidths around 702.2 nm (shown as two circles at
702 nm in Fig.~\ref{fig:tuning}), we just need to consider the cross-sections
of the vertical bars and small area around 702.2 nm. It is then easy to see
that only frequency anti-correlated photon pairs can be detected almost all the
time.

As we use spectral filters with broader bandwidths, the possibility of detecting
uncorrelated photons gets bigger and bigger. It is because when the filter bandwidth is big, uncorrelated photons with large frequency difference
can result in significant accidental coincidence counts because of the apertures used in the experiment. Let us consider an example: the signal photon is at 662 nm  and, due to the energy conservation condition, the idler photon is at 748 nm (shown as two circles at 662 nm and 748 nm). The 662 nm signal photon can be detected.
However, as we can see in Fig.~\ref{fig:tuning}, the 748 nm idler photon simply
cannot be detected because it lies outside the detectable area in the tuning
curve. Therefore, when broadband filters are used, the two-photon wavepacket is mainly determined by spatial filtering by apertures rather than spectral filters. Also, increased non-pair detection or uncorrelated photon detection due the use of broadband filters greatly increases the level of accidental coincidence counts, which in turn reduces the raw quantum interference visibility. 

This effect is clearly demonstrated in Fig.~\ref{fig:dataI}(b) $\sim$
Fig.~\ref{fig:dataI}(d). Up to 20 nm filters, uncorrelated detection evens are
still quite negligible as accidental coincidence counts do not reduce
visibility much. With 80 nm filters, which accepts nearly full bandwidths of
type-I SPDC, the accidental coincidence contribution is huge and this reduces
the raw visibility significantly. Since such accidental coincidence counts from
uncorrelated events produce a flat background, they can be subtracted from the
overall coincidence counts (the corrected visibility increases to 86\%).

It may be possible to remove such spectral filtering effect by removing the apertures altogether or opening them completely. However, this makes it nearly impossible to align and use the interferometer because spatial modes cannot be well defined and the accidental coincidence will increase significantly, just as in this experiment. We have indeed observed slight narrowing of two-photon wavepacket by opening the apertures, but, due to limited collection angles of our detection system, it was not possible to observe very short two-photon wavepacket predicted in section \ref{sec:coh2}. If the bandwidth of type-I SPDC is inherently narrow and the angle-wavelength tuning curve slope is steep, for example, by using a different non-linear crystal or using different phase matching scheme, such an effect may
nevertheless be observed. For example, Burlakov \textit{et. al.} in Ref.~\onlinecite{burlakov} observed a similar effect using collinear type-I SPDC from LiO$_3$ crystals, but the interferometric two-photon wavepacket measurement scheme involved two nonlinear crystals instead of one: i.e., the two-photon wavepacket was measured by interfering two-photon amplitudes from different crystals.

\section{Conclusion}

We have measured one-photon and two-photon wavepackets of type-I and type-II SPDC generated from a cw laser pumped BBO crystal. In the case of type-II SPDC, the measured wavepackets agreed well with the theory. Although we used collinear type-II SPDC for these measurements, it was observed elsewhere that non-collinear type-II SPDC gives the same result  \cite{kg1,kg}. 

In experiments involving type-I SPDC, even though the one-photon wavepacket measurement agreed well with the theory, the two-photon wavepacket was much bigger than the expected value. Upon studying the tuning curve of type-I SPDC from a BBO crystal, we found that spatial filtering limits the two-photon pair detection bandwidth even though one-photon bandwidth is not limited. Such spatial filtering is specific to the tuning curve characteristics of the nonlinear crystal and the geometry of the experiment. It may be avoided with the use of a nonlinear crystal which has sharp (non-collinear) angle-wavelength tuning characteristics or with the use of the collinear multi-crystal geometry \cite{burlakov}.  

In this paper, we have used well-known one-dimensional approximation when calculating one-photon and two-photon wavepackets. This approximation works well with both collinear and non-collinear type-II SPDC experiments with a BBO crystal. However, for broadband non-collinear type-I SPDC, it became clear that tuning curve characteristics of the crystal needs to be considered seriously.

Our results imply that one should be careful using non-collinear type-I SPDC in applications in which variables other than polarizations, such as energy and momentum, are important. One example of such possible applications is quantum metrology; care must be taken not to over-estimate the two-photon bandwidth. Our results also imply that generating high-purity polarization-entangled states or Bell-states using type-I non-collinear SPDC from a BBO crystal almost always rely on strong spectral post-selection, as increased detection bandwidths reduce the raw visibility significantly, even with a cw pump.

\begin{center}
{\bf ACKNOWLEDGEMENTS}
\end{center}

The author wishes to thank M.V. Chekhova for many enlightening discussions and W.P. Grice for helpful comments. This research was supported in part by the U.S. DOE, Office of Basic Energy Sciences, the National Security Agency, and the LDRD program of the Oak Ridge National Laboratory, managed for the U.S. DOE by UT-Battelle, LLC, under contract No.~DE-AC05-00OR22725.


\begin{thebibliography}{}
\bibitem{note} It is possible to completely remove the group velocity dispersion effect from $G^{(2)}(\tau)$ if the positive dispersion introduced in $E_1^{(+)}(t)$ is matched with a negative dispersion introduced in $E_2^{(+)}(t)$, see Ref.~\onlinecite{franson}.
\bibitem{note2} Dispersion cancellation experiment based on this effect is reported in
Ref.~\onlinecite{steinberg} and Ref.~\onlinecite{steinberg2}. This is, however, different from Franson's non-local cancellation of dispersion in $G^{(2)}(\tau)$ described in Ref.~\onlinecite{franson}.
\end{thebibliography}
\end{document}